\documentclass[reprint,amsmath,amssymb]{revtex4-2}
\usepackage[utf8]{inputenc}
\usepackage[T1]{fontenc}
\usepackage{graphicx}
\usepackage[svgnames]{xcolor}
\usepackage{dcolumn}
\usepackage{bm}
\usepackage{enumerate}
\usepackage{hyperref}
\usepackage{mathcomp}
\hypersetup{
 colorlinks=true,%
 linkcolor=MidnightBlue,
 citecolor=MidnightBlue,
 urlcolor=MidnightBlue,
}
\everymath{\displaystyle}
\begin{document}

\preprint{}

\title{Gravity at cosmological distances:\\Explaining the accelerating expansion without dark energy}

\author{Junpei Harada}
 \email{jharada@hoku-iryo-u.ac.jp}
\affiliation{Health Sciences University of Hokkaido, 1757 Kanazawa, Tobetsu-cho, Ishikari-gun, Hokkaido 061-0293, Japan}

\date{August 15, 2023}

\begin{abstract}
Three theoretical criteria for gravitational theories beyond general relativity are considered: obtaining the cosmological constant as an integration constant, deriving the energy conservation law as a consequence of the field equations, rather than assuming it, and not necessarily considering conformally flat metrics as vacuum solutions. Existing theories, including general relativity, do not simultaneously fulfill all three criteria. To address this, a new gravitational field equation is proposed that satisfies these criteria. From this equation, a spherically symmetric exact solution is derived, which is a generalization of the Schwarzschild solution. It incorporates three terms: the Schwarzschild term, the de Sitter term, and a newly discovered term, which is proportional to $r^4$ in a radial coordinate, that becomes significant only at large distances. The equation is further applied to cosmology, deriving an equation for the scale factor. It then presents a solution that describes the transition from decelerating to accelerating expansion in a matter-dominated universe. This is achieved without the need for negative pressure as dark energy or the positive cosmological constant. This provides a novel explanation for the current accelerating expansion of the universe. 
\end{abstract}

\maketitle

\section{Introduction}
In certain gravitational theories beyond general relativity, the cosmological constant $\Lambda$ is derived as a constant of integration. This feature provides a notable theoretical advantage over the Einstein equations in general relativity. Therefore, it is pertinent to establish the following theoretical criteria for gravitational theories:
\begin{enumerate}
	\item The cosmological constant $\Lambda$ is obtained as a constant of integration. 
\end{enumerate}
In the Einstein equations, the presence or absence of the cosmological constant $\Lambda$ is fixed from the beginning. Therefore, general relativity does not meet this criterion. However, the trace-free Einstein equations, denoted by $R_{\mu\nu}-Rg_{\mu\nu}/4=8\pi G (T_{\mu\nu}-Tg_{\mu\nu}/4)$, which were initially investigated by Einstein himself, do satisfy this criterion only if the conservation law, $\nabla_\mu T^\mu{}_\nu=0$, is assumed as an additional assumption~\cite{Ellis:2010uc,Ellis:2013uxa}. The need for this assumption is theoretically a disadvantage. Hence, it is appropriate to require the second theoretical criterion:
\begin{enumerate}
\setcounter{enumi}{1}
	\item The conservation law, $\nabla_\mu T^\mu{}_\nu=0$, is derived as a consequence of the gravitational field equations, rather than being assumed.
\end{enumerate}

General relativity fulfills the second criterion due to the Bianchi identity but fails to satisfy the first criterion. On the other hand, the trace-free Einstein equations fulfill the first criterion but not the second one.

Conformal gravity~\cite{Mannheim:1988dj} and Cotton gravity~\cite{Harada:2021bte,Mantica:2022flg} satisfy both the first and the second criteria. In these theories, the gravitational field equation does not include the cosmological constant; it arises as a constant of integration. Additionally, the conservation law, $\nabla_\mu T^\mu{}_\nu=0$, is derived from the field equations, as in general relativity, due to the Bianchi identity. Unfortunately, in these theories, any conformal flat metric serves as a vacuum solution. This may be a potential disadvantage, as it allows for unphysical solutions. For example, in cosmology, the conformally flat Friedmann-Lemaître-Robertson-Walker metric is a vacuum solution even if the scale factor $a(t)$ is an arbitrary function of $t$, and in this case, it conflicts with observations. Therefore, it is reasonable to consider the third theoretical criterion:
\begin{enumerate}
\setcounter{enumi}{2}
	\item A conformally flat metric is not necessarily a solution in vacuum.
\end{enumerate}

To date, no known theory simultaneously satisfies all three criteria. It remains uncertain whether such a theory is even possible. Hence, the following questions arise: Does a gravitational field equation satisfying the three criteria exist? If so, what is its form? What are the physical implications of such an equation?

This paper provides answers to these questions. First, a new gravitational field equation is proposed, which satisfies all three criteria. Subsequently, from this equation, a spherically symmetric solution is derived, which is a generalization of the Schwarzschild solution. The solution contains three terms: the Schwarzschild term $(\propto 1/r)$, the de Sitter term $(\propto r^2)$, and a newly discovered term $(\propto r^4)$ that only becomes significant at large distances, being negligible at small distances.

The equation is further applied to cosmology. By assuming isotropy and spatial homogeneity of the universe, an equation of motion for the scale factor is derived. The solution to this equation exhibits a significant property: even in the absence of dark energy or the cosmological constant, with only matter present, the universe undergoes a transition from decelerating to accelerating expansion. In fact, in this theory, the accelerating expansion naturally and inevitably emerges as a consequence of the gravitational field equation, rather than being attributed to negative pressure. This offers a novel explanation for the current accelerating expansion of the universe.

This paper is organized as follows. In Sec.~\ref{sec:FE}, we present the gravitational field equation that satisfies the three criteria mentioned earlier. Section~\ref{sec:Schwarzschild} explores a generalized solution of the Schwarzschild solution. In Sec.~\ref{subsec:scalefactor}, we derive the equation of motion for the scale factor, which serves as a generalization of the Friedmann equation. In Sec.~\ref{subsec:accelerating}, we present a solution that describes a transition from decelerating to accelerating expansion in a matter-dominated universe. Finally, Sec.~\ref{sec:summary} provides a summary and conclusions. 

Throughout this paper, we set $c=8\pi G=1$, although $8\pi G$ is explicitly stated in some cases. The covariant derivative uses the Levi-Civita connection, and the metric signature is $(-,+,+,+)$.

\section{Gravitational field equation\label{sec:FE}}
Two different approaches satisfying the first criterion mentioned in the Introduction are known.

The first approach involves demanding that the gravitational field equation be traceless, as originally proposed by Einstein. However, while this approach satisfies the first criterion, it fails to meet the second criterion, thus requiring us to consider an alternative approach.

The second approach employs derivatives of the curvature tensors instead of the curvature tensor itself. This approach includes conformal gravity, Cotton gravity, and Yang's gravitational field equation~\cite{Yang:1974kj}. In a previous study, the author explored a scenario in which the gravitational field equation possesses the same symmetry as $\nabla_\mu R^\mu{}_{\nu\rho\sigma}$. While this approach satisfies the first and the second criteria, it was found to fail to fulfill the third criterion. Therefore, alternative symmetries need to be considered in place of $\nabla_\mu R^\mu{}_{\nu\rho\sigma}$.

Based on these observations, we consider the following scenario. 
We examine two possible totally symmetric derivatives of the curvature:
\begin{subequations}
\begin{eqnarray}
	&\nabla_\rho R_{\mu\nu} + \nabla_\mu R_{\nu\rho} + \nabla_\nu R_{\rho\mu}, \label{eq:1a}\\
	&(g_{\mu\nu} \partial_\rho + g_{\nu\rho} \partial_\mu + g_{\rho\mu} \partial_\nu)R. \label{eq:1b}
\end{eqnarray}
Here, $\nabla_\mu$ represents the covariant derivative, $R_{\mu\nu}$ is the Ricci tensor, and $R$ is the Ricci scalar. These two terms, Eqs.~\eqref{eq:1a} and~\eqref{eq:1b}, are linearly independent, allowing for a linear combination of~\eqref{eq:1a} and~\eqref{eq:1b} to serve as the left-hand side of the gravitational field equation.

A similar representation can be employed for the right-hand side of the gravitational field equation, which comprises two potential terms:
\begin{eqnarray}
	&\nabla_\rho T_{\mu\nu} + \nabla_\mu T_{\nu\rho} + \nabla_\nu T_{\rho\mu}, \label{eq:1c}\\
	&(g_{\mu\nu} \partial_\rho + g_{\nu\rho} \partial_\mu + g_{\rho\mu} \partial_\nu)T. \label{eq:1d}
\end{eqnarray}
\end{subequations}
Here, $T_{\mu\nu}$ is the energy-momentum tensor, and $T$ denotes its trace.

Hence, the gravitational field equation can be expressed as follows:
\begin{eqnarray}
	&a(\nabla_\rho R_{\mu\nu} + \nabla_\mu R_{\nu\rho} + \nabla_\nu R_{\rho\mu})\nonumber\\
	&\qquad+b(g_{\mu\nu} \partial_\rho + g_{\nu\rho} \partial_\mu + g_{\rho\mu} \partial_\nu)R\nonumber\\
	&=c(\nabla_\rho T_{\mu\nu} + \nabla_\mu T_{\nu\rho} + \nabla_\nu T_{\rho\mu})\nonumber\\
	&\qquad+d(g_{\mu\nu} \partial_\rho + g_{\nu\rho} \partial_\mu + g_{\rho\mu} \partial_\nu)T,\label{eq:EOM1}
\end{eqnarray}
where $a$, $b$, $c$, and $d$ are coefficients.

The coefficients $a$, $b$, $c$, and $d$ can be determined as follows. By multiplying Eq.~\eqref{eq:EOM1} by $g^{\nu\rho}$, we obtain
\begin{equation}
	2(a+3b)\partial_\mu R
	=2c \nabla_\lambda T^\lambda{}_{\mu} + (c+6d) \partial_\mu T,
	\label{eq:EOM2}
\end{equation}
where we have used the identity $2\nabla_\mu R^\mu{}_\nu = \partial_\nu R$.

To ensure that the conservation law $\nabla_\mu T^\mu{}_\nu=0$ is satisfied identically, we can derive the following conditions from Eq.~\eqref{eq:EOM2}:
\begin{subequations}
\begin{eqnarray}
	a+3b &=&0,\label{eq:coe1}\\
	c+6d &=&0\label{eq:coe2}.
\end{eqnarray}
\end{subequations}

We also impose the condition that every solution of the Einstein equations satisfies Eq.~\eqref{eq:EOM1}. By substituting $T_{\mu\nu}=R_{\mu\nu}-Rg_{\mu\nu}/2$ (with $8\pi G=1$) and $T=-R$ into the right-hand side of Eq.~\eqref{eq:EOM1}, we obtain
\begin{eqnarray}
	&&(a-c)(\nabla_\rho R_{\mu\nu} + \nabla_\mu R_{\nu\rho} + \nabla_\nu R_{\rho\mu})\nonumber\\
	&&+\left(b+\frac{c}{2}+d\right)(g_{\mu\nu} \partial_\rho + g_{\nu\rho} \partial_\mu + g_{\rho\mu} \partial_\nu)R=0.
\end{eqnarray}
From this equation, we can obtain the following conditions:
\begin{subequations}
\begin{eqnarray}
	a-c &=&0,\label{eq:coe3}\\
	b+\frac{c}{2}+d&=&0\label{eq:coe4}.
\end{eqnarray}
\end{subequations}

Using Eqs.~\eqref{eq:coe1}, \eqref{eq:coe2}, \eqref{eq:coe3}, and~\eqref{eq:coe4} (three of them are linearly independent), we can determine the coefficients $a$, $b$, $c$, and $d$ as follows:
\begin{align}
	&&a=1,
	&&b=-\frac{1}{3},
	&&c=1,
	&&d=-\frac{1}{6},
	\label{}
\end{align}
where we have set $a = 1$ as a normalization.

Here, it is convenient to define the tensor $H_{\mu\nu\rho}$ as
\begin{eqnarray}
	H_{\mu\nu\rho}&\equiv&\nabla_\rho R_{\mu\nu} + \nabla_\mu R_{\nu\rho} + \nabla_\nu R_{\rho\mu}\nonumber\\
	&&-\frac{1}{3}(g_{\mu\nu} \partial_\rho + g_{\nu\rho} \partial_\mu + g_{\rho\mu} \partial_\nu)R,
	\label{eq:def_H}
\end{eqnarray}
which is totally symmetric in $\mu$, $\nu$, and $\rho$. It satisfies
\begin{eqnarray}
	g^{\nu\rho}H_{\mu\nu\rho}=0.
	\label{eq:gH}
\end{eqnarray}
Another convenient definition is
\begin{eqnarray}
	T_{\mu\nu\rho}&\equiv&\nabla_\rho T_{\mu\nu} + \nabla_\mu T_{\nu\rho} + \nabla_\nu T_{\rho\mu}\nonumber\\
	&&-\frac{1}{6}(g_{\mu\nu} \partial_\rho + g_{\nu\rho} \partial_\mu + g_{\rho\mu} \partial_\nu)T,
	\label{eq:def_T}
\end{eqnarray}
which is also totally symmetric in $\mu$, $\nu$, and $\rho$. It satisfies
\begin{eqnarray}
	g^{\nu\rho}T_{\mu\nu\rho}=2\nabla_\nu T^\nu{}_\mu.
	\label{eq:gT}
\end{eqnarray}

Consequently, we obtain the gravitational field equation expressed by third-order totally symmetric tensors,
\begin{equation}
	H_{\mu\nu\rho} = 8\pi G T_{\mu\nu\rho},
	\label{eq:TSFE}
\end{equation}
where we explicitly show $8\pi G=1$.

Multiplying Eq.~\eqref{eq:TSFE} by $g^{\nu\rho}$ and using Eqs.~\eqref{eq:gH} and~\eqref{eq:gT}, we can confirm that the conservation law, $\nabla_\mu  T^\mu{}_\nu=0$, is satisfied as
\begin{equation}
	g^{\nu\rho}H_{\mu\nu\rho} = 16\pi G \nabla_\nu T^\nu{}_\mu =0.
	\label{eq:conservation}
\end{equation}

Here, we provide three remarks on Eq.~\eqref{eq:TSFE}. First, every solution of the Einstein equations satisfies Eq.~\eqref{eq:TSFE}. This means the following: by substituting $8\pi GT_{\mu\nu}=G_{\mu\nu}$ into the right-hand side of Eq.~\eqref{eq:TSFE}, we can confirm that Eq.~\eqref{eq:TSFE} is satisfied. Furthermore, by substituting $8\pi GT_{\mu\nu}=G_{\mu\nu}+\Lambda g_{\mu\nu}$, where $\Lambda$ is nonvanishing, into the right-hand side of Eq.~\eqref{eq:TSFE}, we can confirm that Eq.~\eqref{eq:TSFE} is still satisfied. Thus, Eq.~\eqref{eq:TSFE} does not distinguish between $8\pi GT_{\mu\nu}=G_{\mu\nu}$ and $8\pi GT_{\mu\nu}=G_{\mu\nu}+\Lambda g_{\mu\nu}$. This implies that the cosmological constant $\Lambda$ arises as an integration constant. Second, as shown in Eq.~\eqref{eq:conservation}, the conservation law, $\nabla_\mu T^\mu{}_{\nu}=0$, is satisfied without being assumed. Third, it should be noted that the vanishing of the Weyl tensor $C_{\mu\nu\rho\sigma}$ does not mean the vanishing of $H_{\mu\nu\rho}$. Therefore, a conformally flat spacetime is not necessarily a vacuum solution of $H_{\mu\nu\rho}=0$. Consequently, the gravitational field equation~\eqref{eq:TSFE} simultaneously satisfies all three criteria stated in the Introduction.

\section{Spherically symmetric static vacuum solution\label{sec:Schwarzschild}}

We consider the Schwarzschild-like metric given by 
\begin{equation}
	ds^2 = -e^{\nu (r) } dt^2 + e^{-\nu (r)}dr^2 + r^2 d\Omega^2,
\end{equation}
where $d\Omega^2\equiv d\theta^2+\sin^2\theta d\phi^2$. 
Substituting this into Eq.~\eqref{eq:def_H}, we find that the component of $H_{\mu\nu\rho}$ is expressed as follows:
\begin{align}
	&-2 H^1{}_{11} = -\frac{8}{r^3} \nonumber\\
	&+e^{\nu} \left(\nu^{\prime\prime\prime} + 3 \nu^\prime \nu^{\prime\prime} + (\nu^\prime)^3
	-\frac{2\nu^{\prime\prime}}{r} - \frac{2(\nu^\prime)^2}{r}
	-\frac{2\nu^\prime}{r^2} + \frac{8}{r^3}\right),
	\label{eq:H111}
\end{align}
where a prime denotes the derivative with respect to $r$. The other components vanish except for $H^0{}_{01}$, $H^2{}_{12}$, and $H^3{}_{13}$, which are proportional to $H^1{}_{11}$. By making the substitution
\begin{equation}
	y(r) = \left(\nu^\prime - \frac{2}{r}\right)e^{\nu},
	\label{eq:def_y}	
\end{equation}
Eq.~\eqref{eq:H111} can be simplified to
\begin{equation}
	-2 H^1{}_{11} = y^{\prime\prime} - \frac{6y}{r^2} - \frac{8}{r^3}.
\end{equation}
The solution to the equation $H^{1}{}_{11}=0$ is given by
\begin{equation}
	y(r) = -\frac{2}{r} + \frac{c_1}{r^2} + c_2 r^3,
\end{equation}
where $c_1$ and $c_2$ are constants of integration. Therefore, Eq.~\eqref{eq:def_y} can be rewritten as
\begin{equation}
	\left(\nu^\prime - \frac{2}{r}\right)e^{\nu} = -\frac{2}{r} + \frac{c_1}{r^2} + c_2 r^3.
\end{equation}
This equation can be solved as 
\begin{equation}
	e^\nu = 1 - \frac{c_1}{3r} + c_3 r^2 + \frac{c_2}{2}r^4,
\end{equation}
where $c_3$ is a constant of integration.

If we rename the constants as $c_1=6M$, $c_3 =-\Lambda/3$, and $c_2=-2\lambda/5$, then the solution is given by
\begin{equation}
	-g_{00}=1/g_{11} = e^\nu = 1 - \frac{2M}{r} -\frac{\Lambda}{3} r^2 - \frac{\lambda}{5}r^4.
	\label{eq:SdSH}
\end{equation}
This solution is exact. The term $2M/r$ represents the Schwarzschild term. The term $\Lambda r^2/3$ corresponds to the de Sitter term, indicating that the cosmological constant $\Lambda$ arises as a constant of integration, as expected. The last term, $\lambda r^4/5$, is a newly discovered term that does not emerge from the Einstein equations. When $\lambda$ vanishes (or when $r$ is sufficiently small to ignore the term $\lambda r^4/5$), Eq.~\eqref{eq:SdSH} is the Schwarzschild--de Sitter metric, thus remaining consistent with observations. The term $\lambda r^4/5$ only becomes significant at large distances, such as cosmological distances, and can be ignored at small distances.

We present the curvature invariants as follows:
\begin{subequations}
\begin{eqnarray}
	R^{\mu\nu\rho\sigma}R_{\mu\nu\rho\sigma}  &=&
	\frac{48M^2}{r^6}+\frac{8\Lambda^2}{3}+\frac{48M\lambda}{5r}\nonumber\\
	&&\quad +8\Lambda\lambda r^2+\frac{212\lambda^2 r^4}{25},\\
	R^{\mu\nu}R_{\mu\nu} &=& 4\Lambda^2 + 12 \Lambda \lambda r^2 + 10\lambda^2 r^4,\\
	R &=& 4\Lambda + 6\lambda r^2,\\
	C^{\mu\nu\rho\sigma}C_{\mu\nu\rho\sigma}  &=& \frac{48M^2}{r^6} + \frac{48M\lambda}{5r} + \frac{12 \lambda^2 r^4}{25}.
\end{eqnarray}
\end{subequations}
Hence, we observe that $\lambda$ contributes to both the Ricci tensor and the Weyl tensor. On the other hand, the $M$ only contributes to the Weyl tensor, while the cosmological constant $\Lambda$ contributes solely to the Ricci tensor.

\section{Accelerating universe}
In this section, we apply our gravitational field equation to cosmology. First, we derive the equation of motion for the scale factor. Then, we find a solution that describes the accelerating expansion of the universe.

\subsection{Equation of motion for the scale factor\label{subsec:scalefactor}}
We assume that the universe is isotropic and spatially homogeneous. This assumption leads us to choose a spacetime coordinate system where the metric takes the Friedmann-Lemaître-Robertson-Walker metric~\cite{Friedman:1922kd,Friedmann:1924bb,Lemaitre:1927zz,Lemaitre:1931zza,Lemaitre:1931gd,Robertson:1935zz},
\begin{equation}
	ds^2=-dt^2+a^2(t)\left(\frac{dr^2}{1-kr^2}+r^2 d\Omega^2\right).
	\label{eq:FLRW}
\end{equation}
Here, $a(t)$ represents the scale factor, and $k$ is a constant that represents the curvature of three-dimensional space.
The requirements of isotropy and spatial homogeneity dictate that the components of the energy-momentum tensor take the form
\begin{equation}
	T^\mu{}_\nu = {\rm diag}(-\rho (t), p(t), p(t), p(t)),
	\label{eq:pf}
\end{equation}
and its trace is 
\begin{equation}
	T\equiv T^\mu{}_\mu = -\rho (t) + 3 p(t).
	\label{eq:pf_trace}
\end{equation}
The energy conservation law gives 
\begin{equation}
	0= - \nabla_\mu T^\mu{}_0 = \dot{\rho} + 3 \frac{\dot{a}}{a}(\rho + p),
	\label{eq:conservation_FLRW}
\end{equation}
where $\dot{\rho}\equiv d\rho /dt$ and $\dot{a} \equiv da/dt$. 

We now focus on the gravitational field equation~\eqref{eq:TSFE}. For the Friedmann-Lemaître-Robertson-Walker metric, 
the components of $H_{\mu\nu\rho}$, defined by Eq.~\eqref{eq:def_H}, are given by
\begin{align}
	H^1{}_{01} &= H^2{}_{02} = H^3{}_{03} = -\frac{1}{3}H^0{}_{00}\nonumber\\
	&=-4\left(\frac{\dot{a}}{a}\right)^3 
		- \frac{\dddot{a}}{a}
		+5\frac{\dot{a}\ddot{a}}{a^2}
		-4k\frac{\dot{a}}{a^3}\nonumber\\
	&=\frac{d}{dt}\left[2\left(\frac{\dot{a}}{a}\right)^2 -\frac{\ddot{a}}{a}+\frac{2k}{a^2}\right],
\end{align}
where dots denote time derivatives. The remaining components of $H_{\mu\nu\rho}$ vanish.

We also require the components of the tensor $T_{\mu\nu\rho}$ as defined by Eq.~\eqref{eq:def_T}. From Eqs.~\eqref{eq:FLRW}–\eqref{eq:pf_trace}, we obtain the following expressions:
\begin{equation}
	T^0{}_{00} 
	= -\frac{1}{2} (5\dot{\rho}+3\dot{p})
\end{equation}
and 
\begin{align}
	T^1{}_{01} &= T^2{}_{02} = T^3{}_{03}\nonumber\\
	&=-2\frac{\dot{a}}{a}(\rho+p)+\frac{1}{6}(\dot{\rho}+3\dot{p}).
\end{align}
Using Eq.~\eqref{eq:conservation_FLRW}, we find that 
\begin{align}
	T^1{}_{01} = T^2{}_{02} = T^3{}_{03} = -T^0{}_{00}/3.
\end{align}
The remaining components of $T_{\mu\nu\rho}$ vanish.

The gravitational field equation~\eqref{eq:TSFE} is therefore
\begin{equation}
	\frac{d}{dt}\left[2\left(\frac{\dot{a}}{a}\right)^2 -\frac{\ddot{a}}{a}+\frac{2k}{a^2}\right]
	=8\pi G\frac{d}{dt}\left[\frac{1}{6}(5\rho+3p)\right].
\end{equation}
By integrating this equation, we obtain
\begin{equation}
	2\left(\frac{\dot{a}}{a}\right)^2 -\frac{\ddot{a}}{a} + \frac{2k}{a^2}
	=\frac{4\pi G}{3}(5\rho+3p) + {\rm const},
	\label{eq:Harada1}
\end{equation}
where ${\rm const}$ is a constant of integration. If we rename it as $\Lambda/3$, then we can find that the Friedmann equations
\begin{eqnarray}
	\left(\frac{\dot{a}}{a}\right)^2 &=& \frac{8\pi G}{3} \rho - \frac{k}{a^2} + \frac{\Lambda}{3},
	\label{eq:Friedmann1}\\
	\frac{\ddot{a}}{a} &=& - \frac{4\pi G}{3}(\rho+3p) + \frac{\Lambda}{3},
	\label{eq:Friedmann2}
\end{eqnarray}
satisfy Eq.~\eqref{eq:Harada1}. Thus, in our gravitational theory, the cosmological constant $\Lambda$ is indeed a constant of integration, as expected.

By substituting $\Lambda/3$ for const in Eq.~\eqref{eq:Harada1}, we obtain the following equation for the scale factor:
\begin{equation}
	2\left(\frac{\dot{a}}{a}\right)^2 -\frac{\ddot{a}}{a}
	=\frac{4\pi G}{3}(5\rho+3p) - \frac{2k}{a^2} + \frac{\Lambda}{3}.
	\label{eq:Harada2}
\end{equation}
This is the equation of motion for the scale factor in our gravity theory, which is a generalization of the Friedmann equation in general relativity. It should be noted that the Friedmann equations, Eqs.~\eqref{eq:Friedmann1} and~\eqref{eq:Friedmann2}, with certain $\rho$ and $p$, satisfy Eq.~\eqref{eq:Harada2} with the same $\rho$ and $p$. However, the inverse is not necessarily true; Eq.~\eqref{eq:Harada2}, with certain $\rho$ and $p$, does not necessarily satisfy Eqs.~\eqref{eq:Friedmann1} and~\eqref{eq:Friedmann2} with the same $\rho$ and $p$. Given $p$ as a function of $\rho$, we can solve Eq.~\eqref{eq:conservation_FLRW} to find $\rho$ as a function of $a$. Then, using the obtained $\rho$ as a function of $a$, we can solve Eq.~\eqref{eq:Harada2} to determine $a$ as a function of $t$.

\subsection{Accelerating expansion \label{subsec:accelerating}}
\subsubsection{Preliminary}
In general relativity, Eq.~\eqref{eq:Friedmann2} indicates that the accelerating expansion of the universe $(\ddot{a}>0)$ requires either a positive cosmological constant $\Lambda$ or a negative $\rho + 3p$ (representing dark energy). Therefore, in a matter-dominated universe ($p=0$) with $\Lambda=0$, Eq.~\eqref{eq:Friedmann2} implies that $-\ddot{a}/a  \propto \rho$, indicating decelerating expansion $(\ddot{a}<0)$.

However, in our gravitational theory, the result differs significantly from general relativity. For a matter-dominated universe ($p=0$), with spatial flatness ($k=0$) and $\Lambda=0$, Eq.~\eqref{eq:Harada2} yields $2(\dot{a}/a)^2-\ddot{a}/a\propto \rho$. This does not necessarily imply that $\ddot{a}<0$, because $2(\dot{a}/a)^2$ is positive. In the following, by solving the equation of motion for the scale factor, we will demonstrate the existence of a solution that describes the accelerating expansion in a matter-dominated universe.

\begin{figure*}[thb]
\includegraphics[width=160mm]{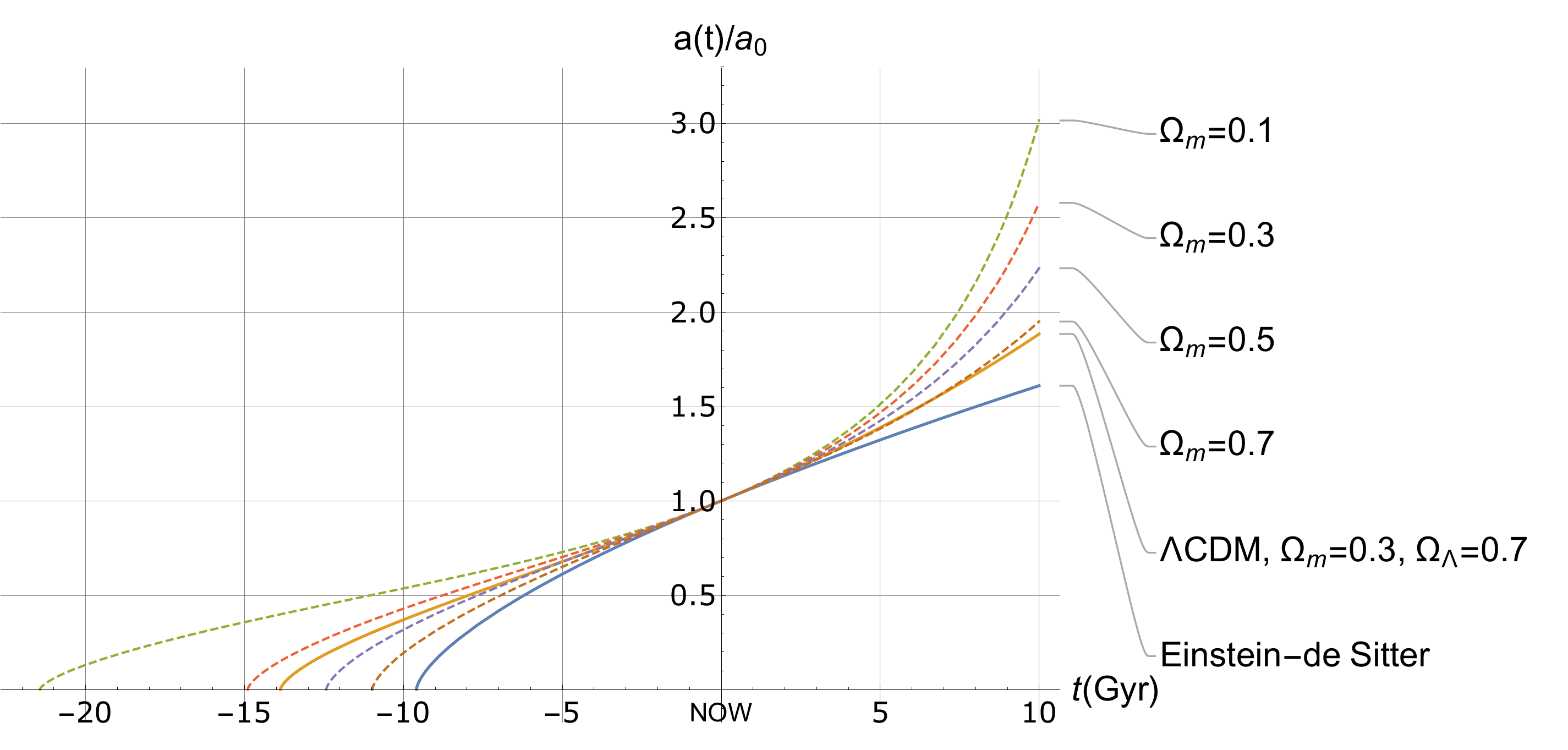}%
\caption{\label{fig:scalefactor}The ratio of the scale factors $a(t)/a_0$ is shown as a function of time $t$ (present is $t=0$) in Gyr. The Hubble constant is assumed to be $H_0=68$ km s$^{-1}$ Mpc$^{-1}$. The four dashed lines represent Eq.~\eqref{eq:scalefactor} for $\Omega_{\rm m}=(0.1, 0.3, 0.5, 0.7)$. These lines clearly demonstrate that even in the absence of dark energy or the cosmological constant, and when only matter is present, the universe undergoes a transition from decelerating to accelerating expansion. The time of this transition, for each case of $\Omega_{\rm m}=(0.1, 0.3, 0.5, 0.7)$, is $-11.4$, $-6.2$, $-3.7$, and $-1.5$ Gyr, respectively. The negative sign of the transition time indicates that the transition occurred in the past $(t<0)$. The age of the universe $t_0$, for each case of $\Omega_{\rm m}=(0.1, 0.3, 0.5, 0.7)$, is $21.4, 14.9, 12.4$, and $11.0$ Gyr, respectively. If we impose the constraint of $t_0>13.0$ Gyr, which is motivated by observations, the range for $\Omega_{\rm m}$ is determined to be $\Omega_{\rm m}<0.44$. The solid blue line represents the Einstein–de Sitter universe with $\Omega_{\rm m}=1$, while the solid orange line represents the $\Lambda$CDM model following the Friedmann equation with $\Omega_{\rm m}=0.3$ and $\Omega_\Lambda=0.7$. The age of the universe $t_0$ is 9.6 Gyr for the Einstein–de Sitter universe, and the age is 13.9 Gyr for the $\Lambda$CDM with $\Omega_{\rm m}=0.3$ and $\Omega_\Lambda=0.7$.}
\end{figure*}

\subsubsection{An accelerating solution}

For simplicity, we assume that the universe is spatially flat $(k=0)$ and $\Lambda=0$. For a matter-dominated universe ($p=0$), Eq.~\eqref{eq:conservation_FLRW} yields
\begin{equation}
 	\rho (t) = \rho_0 \left(\frac{a(t)}{a_0}\right)^{-3},
	\label{eq:matter_density}
\end{equation}
where $a_0$ represents the scale factor at the present time, and $\rho_0$ is the matter density at the present time. Substituting Eq.~\eqref{eq:matter_density}, along with $p=k=\Lambda=0$, into Eq.~\eqref{eq:Harada2}, we obtain
\begin{equation}
	2\left(\frac{\dot{a}}{a}\right)^2 -\frac{\ddot{a}}{a}
	=\frac{5}{2}H_0^2 \Omega_{\rm m} \left(\frac{a}{a_0}\right)^{-3},
	\label{eq:Harada3}
\end{equation}
where $H_0$ is the Hubble constant. The density parameter $\Omega_{\rm m}$, defined as
\begin{equation}
	\Omega_{\rm m} \equiv \frac{\rho_0}{\rho_{\rm c}},\quad
	\rho_c \equiv \frac{3H_0^2}{8\pi G},
\end{equation}
represents the ratio of matter density to the critical density $\rho_{\rm c}$. 
It should be noted that in our gravity theory, $\Omega_{\rm m}$ does not necessarily satisfy $\Omega_{\rm m} + \Omega_{\rm others}=1$, where $\Omega_{\rm others}$ represents the density parameter for other components, if they exist.
Even if $\Omega_{\rm others}=0$, $\Omega_{\rm m}$ is not necessarily equal to 1, because the Friedmann equations are not necessarily satisfied. Therefore, we assume $\Omega_{\rm m}\leq 1$.

The solution to Eq.~\eqref{eq:Harada3} is given by
\begin{align}
	t+t_0 = \frac{2}{3H_0\sqrt{\Omega_{\rm m}}}\left(\frac{a}{a_0}\right)^{3/2}
	{}_2F_1\left(\frac{3}{10},\frac{1}{2};\frac{13}{10};A\left(\frac{a}{a_0}\right)^5\right),
	\label{eq:sol1}
\end{align}
where $t_0$ and $A$ are two constants of integration, and ${}_2F_1\left(a,b;c;x\right)$ is the hypergeometric function. 
The scale factor $a(t)$ reaches zero at $t=-t_0$. By substituting $t=0$ (representing present time) in Eq.~\eqref{eq:sol1}, we obtain
\begin{align}
	t_0 = \frac{2}{3H_0\sqrt{\Omega_{\rm m}}}
	{}_2F_1\left(\frac{3}{10},\frac{1}{2};\frac{13}{10};A\right),
	\label{eq:age1}
\end{align}
where we have used $a_0=a(t=0)$.

We also need to determine the constant $A$. This can be done as follows. 
Differentiating Eq.~\eqref{eq:sol1} with respect to $t$, we have
\begin{align}
	1 = \frac{1}{H_0\sqrt{\Omega_{\rm m}}\sqrt{1-A(a/a_0)^5}}
	\left(\frac{a}{a_0}\right)^{1/2}
	\frac{\dot{a}}{a_0}.
	\label{eq:dif}
\end{align}
Using Eqs.~\eqref{eq:sol1} and~\eqref{eq:dif} to eliminate $1/(H_0\sqrt{\Omega_{\rm m}})$, we find that the Hubble parameter $H(t)\equiv \dot{a}/a$ is given by 
\begin{align}
H(t) = \frac{2\sqrt{1-A(a/a_0)^5}}{3(t+t_0)}
	{}_2F_1\left(\frac{3}{10},\frac{1}{2};\frac{13}{10};A\left(\frac{a}{a_0}\right)^5\right).
	\label{eq:Hubbleparameter}
\end{align}
Therefore, the Hubble constant $H_0 = H(t=0)$ is 
\begin{align}
H_0 = \frac{2\sqrt{1-A}}{3t_0}
	{}_2F_1\left(\frac{3}{10},\frac{1}{2};\frac{13}{10};A\right).
	\label{eq:H0}
\end{align}
By substituting Eq.~\eqref{eq:age1} into Eq.~\eqref{eq:H0}, we obtain
\begin{equation}
	\sqrt{(1-A)\Omega_{\rm m}}=1.
\end{equation}
This yields
\begin{equation}
	A = 1 - \frac{1}{\Omega_{\rm m}} = \frac{\Omega_{\rm m}-1}{\Omega_{\rm m}}.
	\label{eq:constantA}
\end{equation}

Taking the ratio between Eqs.~\eqref{eq:sol1} and~\eqref{eq:age1} and using Eq.~\eqref{eq:constantA}, we find that the scale factor $a(t)$ satisfies 
\begin{align}
	\frac{t+t_0}{t_0} 
	= \frac{{}_2F_1\left(\frac{3}{10},\frac{1}{2};\frac{13}{10};\frac{\Omega_{\rm m}-1}{\Omega_{\rm m}}\left(\frac{a}{a_0}\right)^5\right)}{{}_2F_1\left(\frac{3}{10},\frac{1}{2};\frac{13}{10};\frac{\Omega_{\rm m}-1}{\Omega_{\rm m}}\right)}
	\left(\frac{a}{a_0}\right)^{3/2}.
	\label{eq:scalefactor}
\end{align}
Here, the age of the universe $t_0$ is given by
\begin{align}
t_0 = \frac{2}{3H_0\sqrt{\Omega_{\rm m}}}
	{}_2F_1\left(\frac{3}{10},\frac{1}{2};\frac{13}{10};\frac{\Omega_{\rm m}-1}{\Omega_{\rm m}}\right).
	\label{eq:age2}
\end{align}
These equations, Eqs.~\eqref{eq:scalefactor} and~\eqref{eq:age2}, are fundamental equations.
Using the observed value of the Hubble constant $H_0$, Eq.~\eqref{eq:age2} determines $t_0$ as a function of $\Omega_{\rm m}$.
Subsequently, using $t_0$ as a function of $\Omega_{\rm m}$, Eq.~\eqref{eq:scalefactor} determines $a(t)$ as a function of $t$ and $\Omega_{\rm m}$.

In the special case where $\Omega_{\rm m}=1$, Eqs.~\eqref{eq:scalefactor} and~\eqref{eq:age2} simplify to
\begin{align}
	&&\frac{a}{a_0} = \left(\frac{t+t_0}{t_0}\right)^{2/3},
	&&t_0 =\frac{2}{3H_0}.
\end{align}
These equations represent the Einstein–de Sitter universe, and thus Eqs.~\eqref{eq:scalefactor} and~\eqref{eq:age2} include the result derived from general relativity as a special case. 

In general cases where $\Omega_{\rm m}\not=1$, Fig.~\ref{fig:scalefactor} illustrates the behavior of $a(t)/a_0$ as a function of time. The figure demonstrates that even in the absence of dark energy or the cosmological constant, and with only matter present, the universe undergoes a transition from decelerating to accelerating expansion.

This transition, from deceleration to acceleration, occurs at $t=-t_\star$, which is the time when the acceleration $\ddot{a}$ reaches zero.
By performing a straightforward calculation, we can determine the time $t_\star$ as follows:
\begin{equation}
	t_\star = t_0 \left[1 - \left(\frac{\Omega_{\rm m}}{4(1-\Omega_{\rm m})}\right)^{3/10} 
	\frac{{}_2F_1\left(\frac{3}{10},\frac{1}{2};\frac{13}{10};-\frac{1}{4}\right)}{{}_2F_1\left(\frac{3}{10},\frac{1}{2};\frac{13}{10};\frac{\Omega_{\rm m}-1}{\Omega_{\rm m}}\right)}\right],
\end{equation}
where ${}_2F_1 (3/10,1/2;13/10;-1/4)\approx 0.97383$. 
This equation indicates that $t_\star$ is positive if $\Omega_{\rm m}$ is less than 0.8. 
A positive $t_\star$ (or equivalently negative $-t_\star$) implies that the transition from deceleration to acceleration occurred in the past ($t<0$). Therefore, we can conclude that in our gravitational theory, 
even in the absence of dark energy,
the current universe is in an accelerating phase if $\Omega_{\rm m}<0.8$. The transition time ($t=-t_\star$) for typical values of $\Omega_{\rm m}$ is provided in the caption of Fig.~\ref{fig:scalefactor}.

\section{Summary and conclusions\label{sec:summary}}
In this study, we have set three theoretical criteria for gravitational theories, as outlined in the Introduction:
\begin{enumerate}
	\item The gravitational field equations should not explicitly contain the cosmological constant $\Lambda$, but it can emerge as a constant of integration.
	\item The conservation law $\nabla_\mu T^\mu{}_{\nu}=0$ should be derived as a consequence of the field equations, rather than being introduced as an additional assumption.
	\item A conformally flat metric should not necessarily be a vacuum solution.
\end{enumerate}
These criteria impose stringent restrictions on gravitational theories, and so far, no theory has been known to fulfill all three criteria. In this paper, we have presented the gravitational field equation, Eq.~\eqref{eq:TSFE}, which satisfies all three criteria. Our construction provides an explicit model, and while it is a unique model that the author could find, it may not be the only one. These criteria and their fulfillment are summarized in Table.~\ref{tab:table1}.

\begin{table}[tbh]%
\caption{\label{tab:table1}%
A summary of typical gravitational theories and their fulfillment against the three criteria.
}
\begin{ruledtabular}
\begin{tabular}{lccccc}
\textrm{Criterion}&
\textrm{GR\footnote{General relativity.}}&
\textrm{TFE\footnote{Trace-free Einstein equations.}}&
\textrm{CG\footnote{Conformal gravity.}}&
\textrm{Cotton\footnote{Cotton gravity.}}&
\textrm{This work}
\\
\colrule
1 & No & Yes & Yes & Yes & Yes\\
2 & Yes & No & Yes & Yes & Yes\\
3 & Yes & Yes & No & No  & Yes\\
\end{tabular}
\end{ruledtabular}
\end{table}

Additionally, we have derived a spherically symmetric solution that generalizes the Schwarzschild solution. This solution consists of three terms: the Schwarzschild term $(\propto 1/r)$, the de Sitter term $(\propto r^2)$, and a newly discovered term $(\propto r^4)$. The $r^4$ term only becomes significant at large distances while being negligible at short distances. This indicates that gravity described by Eq.~\eqref{eq:TSFE} differs from general relativity primarily at large distances, such as cosmological distances.

Motivated by this observation, we have applied our gravitational equations to cosmology. Assuming the isotropy and spatial homogeneity of the universe, we have derived an equation for the scale factor, Eq.~\eqref{eq:Harada2}, which serves as a generalization of the Friedmann equation. Through our analysis, we have demonstrated that even in the absence of dark energy or the cosmological constant,  the universe undergoes a transition from a decelerating phase to an accelerating phase. Thus, in our gravitational theory, the current accelerating expansion is a natural and inevitable consequence in a matter-dominated universe.

\begin{acknowledgments}
This work was supported by JSPS KAKENHI Grant No. JP22K03599.
\end{acknowledgments}

\bibliography{references}
\end{document}